\documentclass[preprint,preprintnumbers,amsmath,amssymb,floatfix,showpacs,aps]{revtex4}
\usepackage{epsfig}

\begin{document}
\title{Intermediate-energy electron-impact dissociative ionization-excitation of molecular hydrogen}
\author{Vladislav V. Serov}
\affiliation{Department of Theoretical Physics, Saratov State University, 83 Astrakhanskaya, Saratov 410012, Russia}
\author{Boghos B. Joulakian}
\affiliation{Universit\'e de Lorraine, SRSMC(UMR CNRS 7565), 1 boulevard Arago, bat. ICPM 57078 Metz Cedex 3, France}

\date{\today}

\begin{abstract}

We have implemented three variants of the exterior complex scaling
procedure in prolate spheroidal coordinates (PS-ECS) to study the
dissociative electron impact ionization-excitation of hydrogen
molecule, where the emerging electrons and one of the protons are
detected in coincidence for the first time in a recent experiment.
In the first variant, designated PSECS-1B, the two target electrons
are treated \textit{ab initio} while the interaction of the
incident-scattered electron is taken into account using the first
term of the Born series. In the second, PSECS-2BCD, the second Born
term is introduced in the dipole approximation. In the third
approach, designated PSECS-SW, applied to the ionization-excitation
to the $2p\sigma_u$ level of H$_2^+$, the multi-configurational
single active electron approximation is used for the target, while
the interaction of the incident electron with the target is described
\textit{ab initio}. Our results agree partially with those of a
recent experiment which is in progress.
\end{abstract}


\maketitle

\section{Introduction}

Simple (e,2e) ionization of atomic and molecular targets
designates complete inelastic electron-target collision
experiments, where the scattered and the ejected electrons from
the target are detected in coincidence
\cite{Weigold76,Ehrhardt86,Lahmam,Whelan93}. In the case of
diatomic molecules, one has to separate the vibrational and
rotational movements (see for that \cite{Ijima}) and average over
all possible directions of the molecule in the laboratory frame
with respect to which the emerging electrons are detected.

In recent years many molecular orientation resolved experiments
have been undertaken
\cite{Takahashi,Bellm,Senftleben,Senftleben2}. The main
experimental methods to realize these experiments are either by
aligning the target before the collision by exciting its
rotational movement by a polarized laser \cite{Rosca} or/and by
detecting the emerging dissociated nuclei in coincidence with the
ejected and scattered electrons. Recently, the latter approach
was applied to the simple ionization excitation of molecular
hydrogen H$_2$ to the $2s\sigma_g$, $2p\pi_u$ and $2p\sigma_u$
levels of H$_2^+$ \cite{Bellm}. Compared to the situation in the
COLTRIMS method \cite{Senftleben}, the emerging proton is detected
here with higher energy resolution, which represents a quite
important step in this domain. We will designate this type of
coincidence experiments (e,2e+p) simple ionization.

We have in the past determined the multiply differential cross
section of this process \cite{Lahmidi,Serov2005,Joulakian} for
high (5 keV) incident energy values and have observed interesting
interference phenomena and have shown like in
\cite{Senftleben,Senftleben2} that the proton emerges
preferentially in the direction of the momentum vector transferred
to the system by the incident electron \cite{Joulakian,Serov2011}.

In the present paper, we apply the exterior complex scaling method
in prolate spheroidal coordinates (PS-ECS) \cite{Serov2009} to the
determination of the multi-fold differential cross section MDCS of
the ionization-excitation of H$_2$ to the $2s\sigma_g$, $2p\pi_u$
and $2p\sigma_u$ levels of H$_2^+$. Our aim is to gain physical
insight in this complex problem, to interpret the results of
\cite{Bellm} and give theoretical guidance to near future
experiments.

\section{Theory}\label{Theory_section}

We apply three variants based on the exterior complex scaling
method in prolate-spheroidal coordinates (PS-ECS) developed in
\cite{Serov2009}. This method is based on the solution of the
six-dimensional driven Schr\"{o}dinger equation in prolate
spheroidal (elliptic) coordinates
\begin{eqnarray}
(\hat{H}-E)\psi^{(+)} = -\hat\mu\psi_{i}\label{drivenSchr}
\end{eqnarray}
where $\psi^{(+)}$ is the perturbed part of the wave function,
with the boundary conditions of an out-going wave provided by the
exterior complex scaling (ECS). The $E$ is the final energy of the
electrons are described by $\psi^{(+)}$.

The scattering of the electron on two-electron molecule results in
a nine-dimensional Schr\"{o}dinger equation, which we reduce first
to a six-dimensional equation. The nuclei are considered to be
fixed during the process. The choice of the appropriate approach
to do this reduction depends on the mechanism that we want to
study. In the ionization-excitation, one of the target electrons
can be ejected by the incident electron, while the second can be
excited by a second impact (sequential mechanism). We can also
have the excitation of the second electron by the inter-electron
correlations in the initial target state (kick-off mechanism), or
in the final target state (final-state scattering).

For a relatively fast incident electron, the natural approach is
the application of the Born series for the incident and scattered
electron. In this approach,
 $\psi^{(+)}(\mathbf{r}_1,\mathbf{r}_2)$
represents the wave function of the (unknown) final state of the
target electrons with coordinates $\mathbf{r}_1$ and
$\mathbf{r}_2$ in Eq.(\ref{drivenSchr}). The $\hat{H}$ is the
Hamiltonian of the target. The $E$ designates the final energy of
the target electrons and $\psi_{i}(\mathbf{r}_1,\mathbf{r}_2)$ the
wave function of the initial state of the target. The excitation
operator $\hat\mu$ takes into account the effect of the incident
electron.

The first order Born term of the excitation operator has the form
\cite{Serov2009}
\begin{eqnarray}
\hat{\mu}_{1B}&=&-\frac{1}{2\pi}\langle\mathbf{k}_s|V|\mathbf{k}_i\rangle=
-\frac{2}{K^2}\left[e^{i\mathbf{K}\cdot\mathbf{r}_1}+e^{i\mathbf{K}\cdot\mathbf{r}_2}-e^{i\mathbf{K}\cdot\mathbf{R}/2}-e^{-i\mathbf{K}\cdot\mathbf{R}/2}\right],
\label{FBexp}
\end{eqnarray}
where $R$ is internuclear distance, and $\mathbf{R}=R\mathbf{n}_R$
is the vector with the orientation coinciding with the molecular
axis orientation. Next, $V$ is the potential energy term of the
interaction of the molecule with the incident electron,
$\mathbf{k}_i$ is its momentum, $\mathbf{k}_s$ represents the
momentum of the scattered electron, and $\mathbf{K}$ the momentum
transfer. Here we assume
$|\mathbf{k}\rangle=\exp(i\mathbf{k}\cdot\mathbf{r}_0)$,
$\mathbf{r}_0$ being the position vector of the fast
incident/scattered electron. With this choice, the kick-off and
the final-state scattering processes are taken into account, but
not the sequential mechanism. This will be done by considering the
second Born effects in a relatively simplified manner
\cite{Serov2010}. In fact, we add to the excitation operator
$\hat\mu$ the corrected double dipole approximation of the second
Born term \cite{Serov2010}
\begin{eqnarray}
\hat{\mu}_{2BDC}(\mathbf{r}_1,\mathbf{r}_2)=\sum_{M_1,M_2=-1}^{1}\mathcal{M}^{M_1M_2}(x_{1M_1}+x_{2M_1})(x_{1M_2}+x_{2M_2}).\label{dipole2B}
\end{eqnarray}
Here $\mathcal{M}^{M_1M_2}$ is the second-order cross derivative
term of the Born series for the scattered electron in the closure
approximation, $x_{1M}$ and $x_{2M}$ are components of the vectors
$\mathbf{r}_1$ and $\mathbf{r}_2$, respectively. The two dipole
factors act on both target electrons.

The Born series approach is not well adapted to our problem, where
we have an intermediate electron impact energy of 178 eV, and the
plane wave description of the incident-scattered electron is not
quite appropriate. So we implement an alternative approach, in
which the interaction of the incident electron with the target is
described precisely, but the interaction between the target
electrons in the final state is omitted. The initial state of
$H_2$ can be expanded over products of bound states of $H_2^+$
\begin{eqnarray}
\Phi_0(\mathbf{r}_1,\mathbf{r}_2)=\sum_{n_1,l_1,m,n_2,l_2}c_{n_1l_1n_2l_2m}\varphi_{n_1l_1m}(\mathbf{r}_1)\varphi_{n_2l_2\,-m}(\mathbf{r}_2),\label{expansionPhi0}
\end{eqnarray}
where $n,l,m$ is the set of spheroidal quantum numbers, which
specify the bound states $\varphi_{nlm}(\mathbf{r})$ of H$_2^+$,
$c_{n_1l_1n_2l_2m}$ are the expansion coefficients.
Each of the terms in the sum \eqref{expansionPhi0} can be
considered as an electron configuration. Under the assumption,
that the ejection of one of the electrons does not change the
other electron state, and the final ion state is $\varphi_{n_2l_2\,-m}$, the initial state of the electron to be ejected may be described by the function
\begin{eqnarray}
\varphi_0(\mathbf{r}_1)=\langle \varphi_{n_2l_2\,-m}(\mathbf{r}_2)| \Phi_0(\mathbf{r}_1,\mathbf{r}_2)\rangle.\label{projPhi0}
\end{eqnarray}
Then, following \eqref{expansionPhi0}, the initial state of the electron to be ejected may be expressed as 
\begin{eqnarray}
\varphi_0(\mathbf{r}_1)=\sum_{n,l}c_{nln_2l_2m}\varphi_{nlm}(\mathbf{r}_1).\label{projPhi0exp}
\end{eqnarray}
Each term of this expansion can be associated with the unperturbed wave function, describing a system ``bound electron + incoming fast electron''
\begin{eqnarray}
\psi_{i}(\mathbf{r}_0,\mathbf{r}_1)=\frac{1}{\sqrt{2}}\left[\exp(i\mathbf{k}_i\cdot\mathbf{r}_0)\varphi_{nlm}(\mathbf{r}_1)+(-1)^S\exp(i\mathbf{k}_i\cdot\mathbf{r}_1)\varphi_{nlm}(\mathbf{r}_0)\right],
\end{eqnarray}
where $S$ is the total spin of the incident and target
electrons possessing the value $S=0$ with probability 1/4, or the value $S=1$ with probability 3/4. This function may be substituted in right-hand-side of Eq.\eqref{drivenSchr} allowing to obtain the wave function
$\psi^{(+)}(\mathbf{r}_0,\mathbf{r}_1)$ of the scattered
and the ejected electrons. For this aim in Eq.\eqref{drivenSchr} $E$ should be set equal to the total energy of the incident
and the target electron, and the operator $\hat\mu$ equal to the potential of the interaction between the incident and the target electrons
\begin{eqnarray}
\hat\mu=\frac{1}{|\mathbf{r}_0-\mathbf{r}_1|}.
\end{eqnarray}
In turn, from the scattering wave function $\psi^{(+)}(\mathbf{r}_0,\mathbf{r}_1)$ one may extract the scattering amplitude \cite{ReviewMcCurdy2004}. The latter is in fact the amplitude of the impact ionization of H$_2^+$ initially being in $\varphi_{nlm}$ state. Since the electron spin directions are unchanged during the scattering, the total cross section $\sigma_{\text{H}_2^+}(nlm)$ of H$_2^+[nlm]$(e,2e) might be combined from the corresponding partial cross sections for both values of $S$ with the respective weight factors.
 
Finally, reminding the Eq.\eqref{projPhi0exp} we arrive to the expression of the H$_2$ impact ionization-excitation cross section via the sum of the cross sections $\sigma_{\text{H}_2^+}(n_1l_1m_2)$ of the impact ionization of the H$_2^+$ in the initial $\varphi_{nlm}$ states: 
\begin{eqnarray}
\sigma_{n_2l_2m_2}=\sum_{n_1,l_1}|c_{n_1l_1n_2l_2m_2}|^2\sigma_{\text{H}_2^+}(n_1l_1m_2). \label{sumH2plus_channels}
\end{eqnarray}
In what follows, we will refer to this approach as
PSECS-SW (SW for scattered wave). In this approximation the inter-electron correlations are taken into account only in the H$_2$ initial state, while neglecting the interaction between the molecular electrons during the impact. It means, that we consider only the kick-off process, while the final-state scattering and the sequential mechanism are omitted. On the other hand, the distortion of the incoming wave due to the interaction with the nuclei and the post-collisional interaction of the ejected electron with the scattered electron is taken into account. 

The comparison of the results of PSECS-1B, PSECS-2B and PSECS-SW on the ionization-excitation of H$_2$ to the $2p\sigma_u$ level of H$_2^+$ with the experimental data should reveal the dominant effects in each experimental situation.

\section{Results}\label{Results_section}

In this part we will compare our results to the experimental ones
published in \cite{Bellm}.  Some supplementary experimental
results of the same authors are also available on the site of the
Many Particle Spectroscopy Conference held in Berlin in 2012
\cite{MPS}. Perpendicular and parallel alignments of the
internuclear axis to the incidence direction and to the momentum
transfer directions are considered. In all these cases the
variation of the MDCS of (e,2e) simple ionization and of H$_2$
with formation of the residual H$_2^+$ in $2s\sigma_g$, $2p\pi_u$,
$2p\sigma_u$ states are given in terms of the scattering angle
$\theta_s$.

Now, the most common feature of all (e,2e) simple ionization
experiments is that the binary peek, which shows the preferential
ejection direction, is oriented around that of the momentum
transfer. In the experimental conditions here, this region is
found around $35^\circ\leq\theta_e\leq65^\circ$ measured with
respect to the incidence direction (all electrons are detected in
the incidence plane, the $\theta_s$ and $\theta_e$ are measured
with respect to the direction of $\mathbf{k}_i$).  Following
\cite{Bellm}, we have thus integrated the MDCS over the ejection
angle $\theta_e$ of the slow electron.
\begin{eqnarray}
\frac{d^3\sigma}{dE_e d\phi_e  d\Omega_s}=\int_{\theta_a}^{\theta_b}\frac{d^3\sigma}{dE_e d\Omega_e  d\Omega_s}d\theta_e.
\end{eqnarray}

All our PSECS-1B/2B calculations were performed for the total
electronic energy value $E=23.3$ eV which represents the sum of
the energy values of the ejected electron and the binding energy
of the residual ion $E=E_e+E_{\text{ion}}$. The probability
amplitudes of the ionization-excitation with the formation of the
residual ion in the $2s\sigma_g$, $2p\pi_u$ and $2p\sigma_u$
states were extracted from a single wave function
$\psi^{(+)}(\mathbf{r}_1,\mathbf{r}_2)$. Due to the fixed total energy $E$, in our calculations the ejected electron energies $E_e=E-E_{\text{ion}}$ were different for the different final ion states: $40$ eV (that correspond to its value in the experiment \cite{Bellm,MPS}) in the case of $2p\sigma_u$ final ion state, $35.8$ eV in the case of $2p\pi_u$ and $34$ eV in the case of $2s\sigma_g$ state. Since the difference between the ionization energy values of H$_2$ to the $2s\sigma_g$ and $2p\pi_u$ level of H$_2^+$ is smaller than the energy resolution of the experiment \cite{Bellm,MPS}, we have summed the corresponding MDCS for these two levels. So, the average ejected electron energy was appeared to be equal to $35$ eV for the $2s\sigma_g+2p\pi_u$ case.

\begin{figure}[ht]
\includegraphics[angle=-90,width=0.45\textwidth]{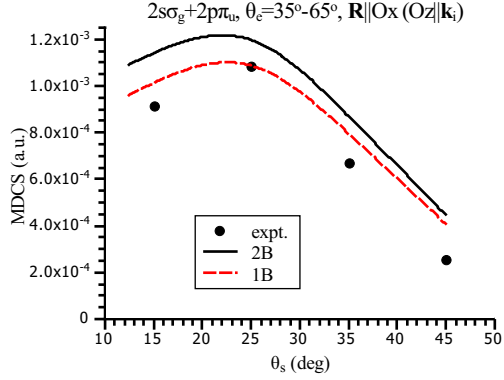}
\\(a)\\
\includegraphics[angle=-90,width=0.45\textwidth]{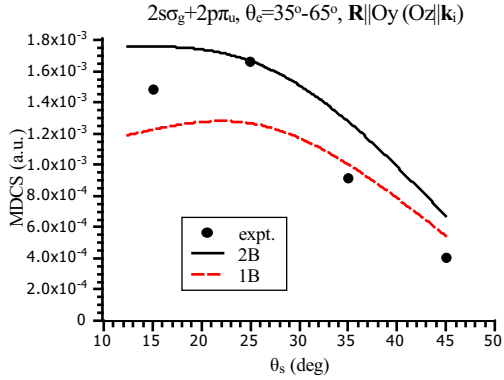}
\\(b)\\
\includegraphics[angle=-90,width=0.45\textwidth]{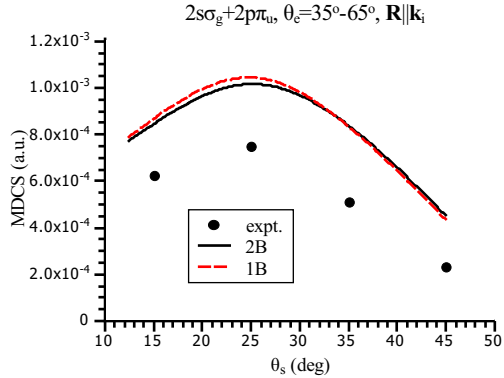}
\\(c)\\
\caption{(Color online) The variation of the MDCS of (e,2e) simple
ionization of H$_2$ with formation the residual H$_2^+$ in
$2s\sigma_g$ and $2p\pi_u$ states, integrated over the ejection
angle of slow electron $\theta_e$ between
$35^\circ\leq\theta_e\leq65^\circ$, in terms of the scattering
angle $\theta_s$. The energy of the incident electron $E_i=179$eV,
and that of the ejected electron $E_e=35$ eV. Results of PSECS-2B
are presented by the solid line, those of PSECS-1B by the dashed
line and the experimental data \cite{Bellm} by dots. The
orientation of the internuclear axis : a) $\mathbf{R}\parallel
Ox$, b) $\mathbf{R}\parallel Oy$, c) $\mathbf{R}\parallel Oz$.
$Oz\parallel \mathbf{k}_i$. \label{FIGsigma2ssg}}
\end{figure}

\begin{figure}[ht]
\includegraphics[angle=-90,width=0.45\textwidth]{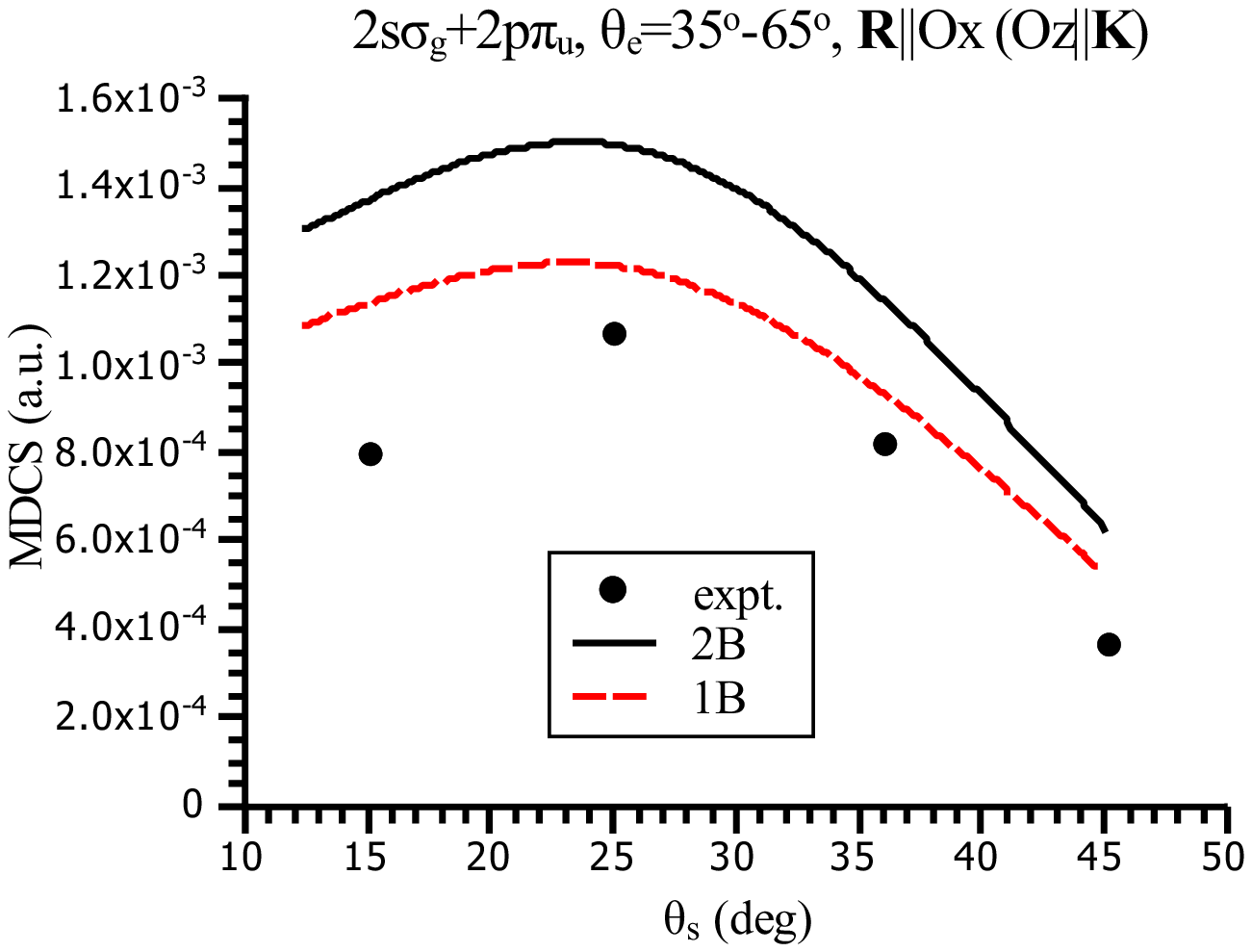}
\\(a)\\
\includegraphics[angle=-90,width=0.45\textwidth]{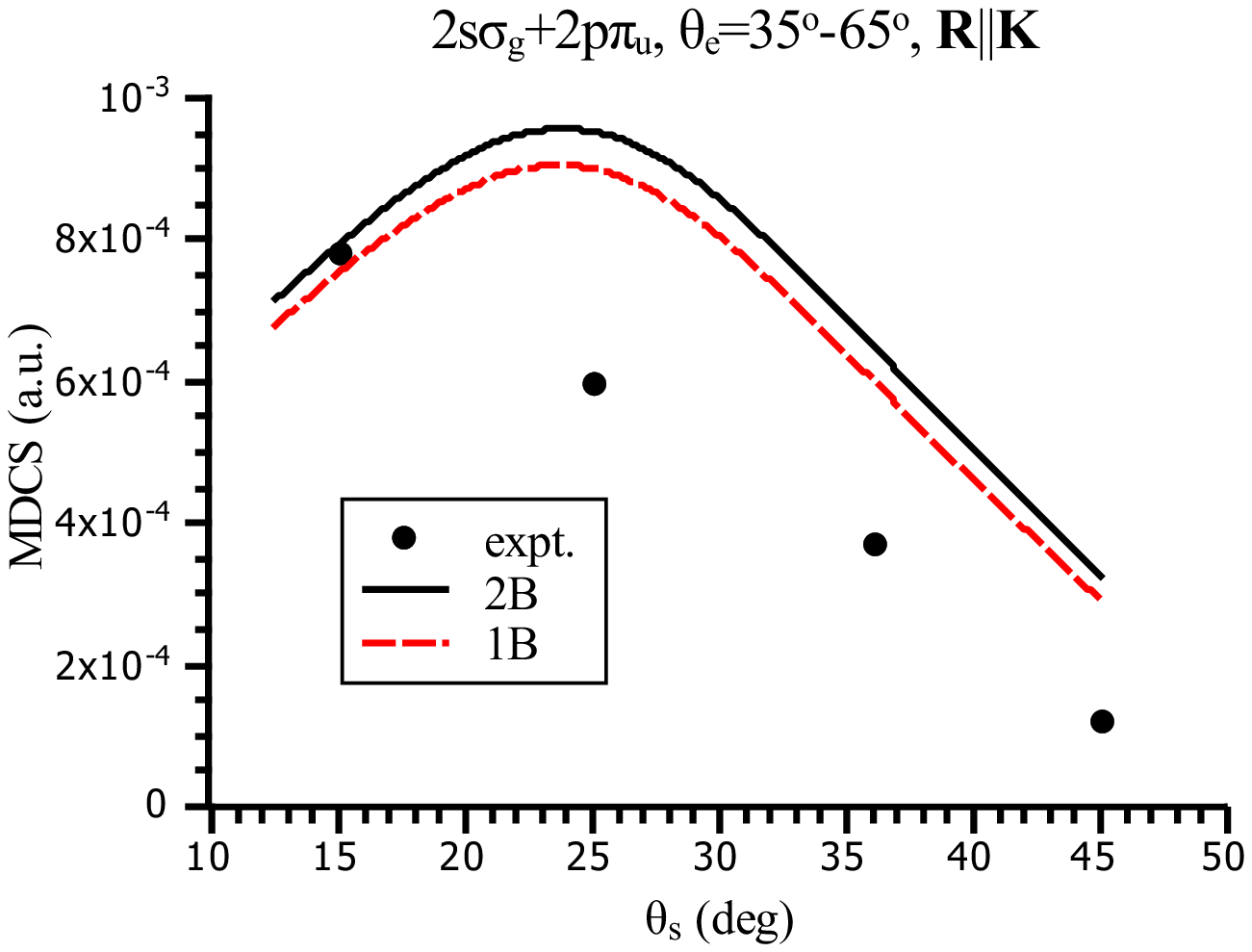}
\\(b)\\
\caption{(Color online) The same as on Fig.\ref{FIGsigma2ssg}, but
for $Oz \| \mathbf{K}$. In a) $\mathbf{R}\parallel Ox$; b)
$\mathbf{R}\parallel Oz$.
The case $\mathbf{R}\parallel Oy$ is already presented in the Fig.\ref{FIGsigma2ssg}(b).
\label{FIGsigma2ssgK}}
\end{figure}

On figures \ref{FIGsigma2ssg} and \ref{FIGsigma2ssgK} we compare
our results for ionization-excitation H$_2$ with formation the
residual H$_2^+$ in $2s\sigma_g$ and $2p\pi_u$ states to the
experimental ones for the above mentioned orientations of the
molecule. In spite of the fact, that the simple plane wave
description of the incident and scattered electrons is used in
PSECS-1B, the results show the same structure as the experimental
points. We observe also, that the introduction of the second Born
terms in PSECS-2B does not bring any change in the structure of
the graphs. This can be explained by the fact that kick-off and
final-state scattering are more important than the sequential
mechanism in these situations. One can also observe on figure
\ref{FIGsigma2ssgK}b, which corresponds to the situation where the
molecule is aligned with the direction of the momentum transfer,
the absence of the maximum around $\theta_s=25^\circ$, in contrast
to the four preceding cases shown on Figures \ref{FIGsigma2ssg}
and \ref{FIGsigma2ssgK}.

\begin{figure}[ht]
\includegraphics[angle=-90,width=0.45\textwidth]{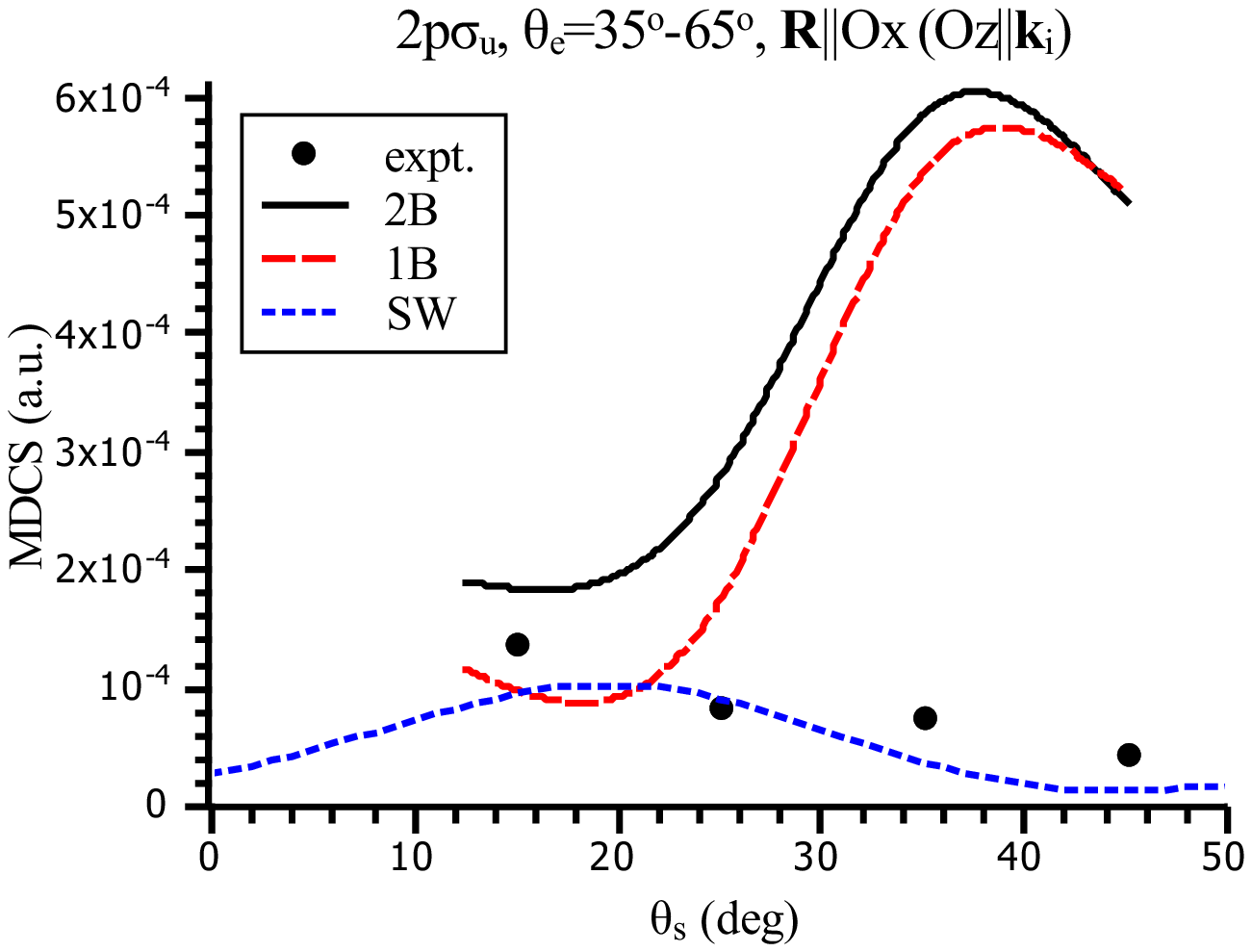}
\\(a)\\
\includegraphics[angle=-90,width=0.45\textwidth]{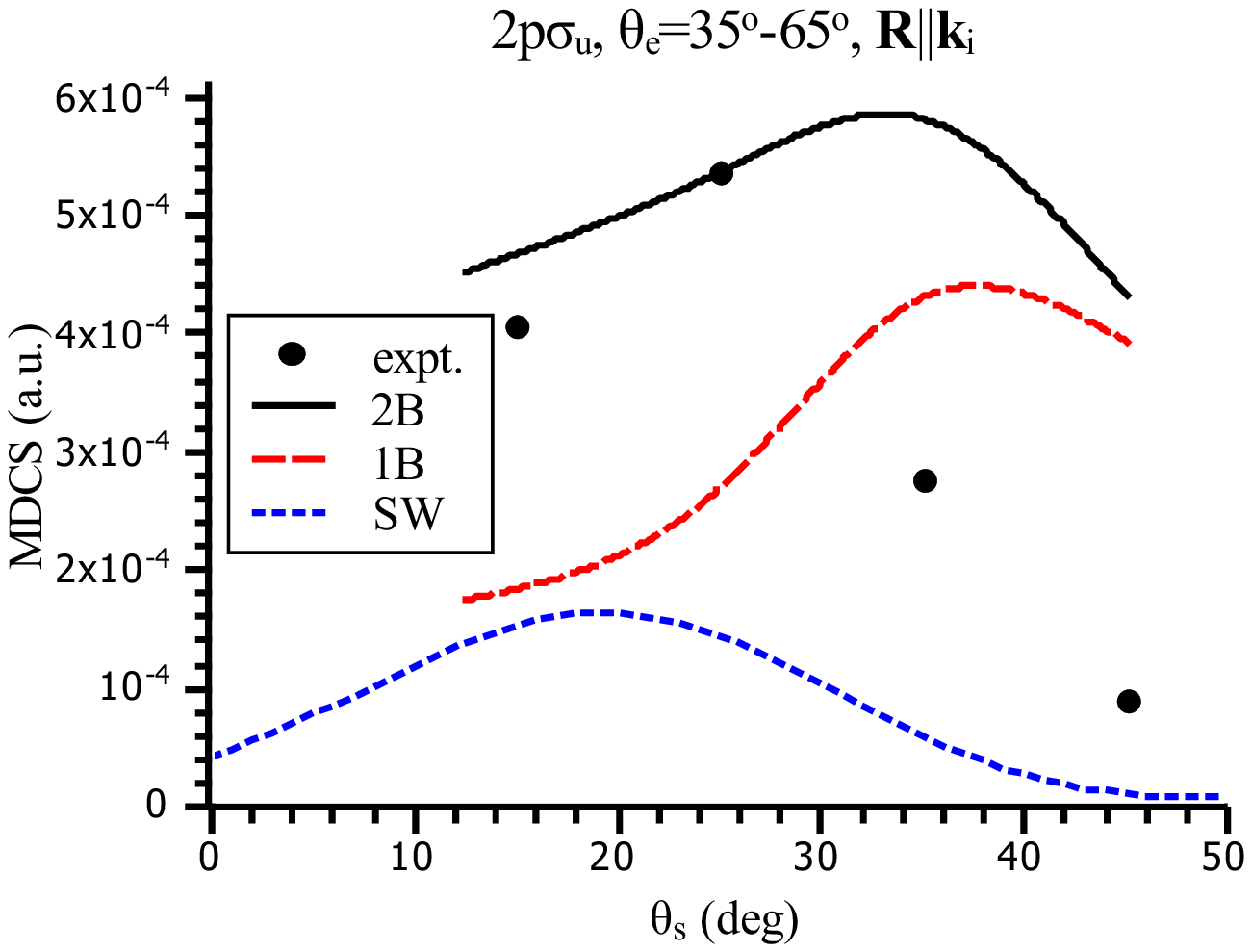}
\\(b)\\
\caption{(Color online) The same as on Fig.\ref{FIGsigma2ssg}, but
with formation of residual ion in the $2p\sigma_u$ level. Here the
ejection energy $E_e=40$ eV. Results of PSECS-2B (solid line),
PSECS-1B (dashed line), PSECS-SW (dotted line), experimental data
\cite{Bellm} (dots).  a) $\mathbf{R}\parallel Ox$; b)
$\mathbf{R}\parallel Oz$. The experimental results concerning the
case $\mathbf{R}\parallel Oy$ are not shown here since they are
relatively small with respect to the other cases.
\label{FIGsigma2psu}}
\end{figure}

\begin{figure}[ht]
\includegraphics[angle=-90,width=0.45\textwidth]{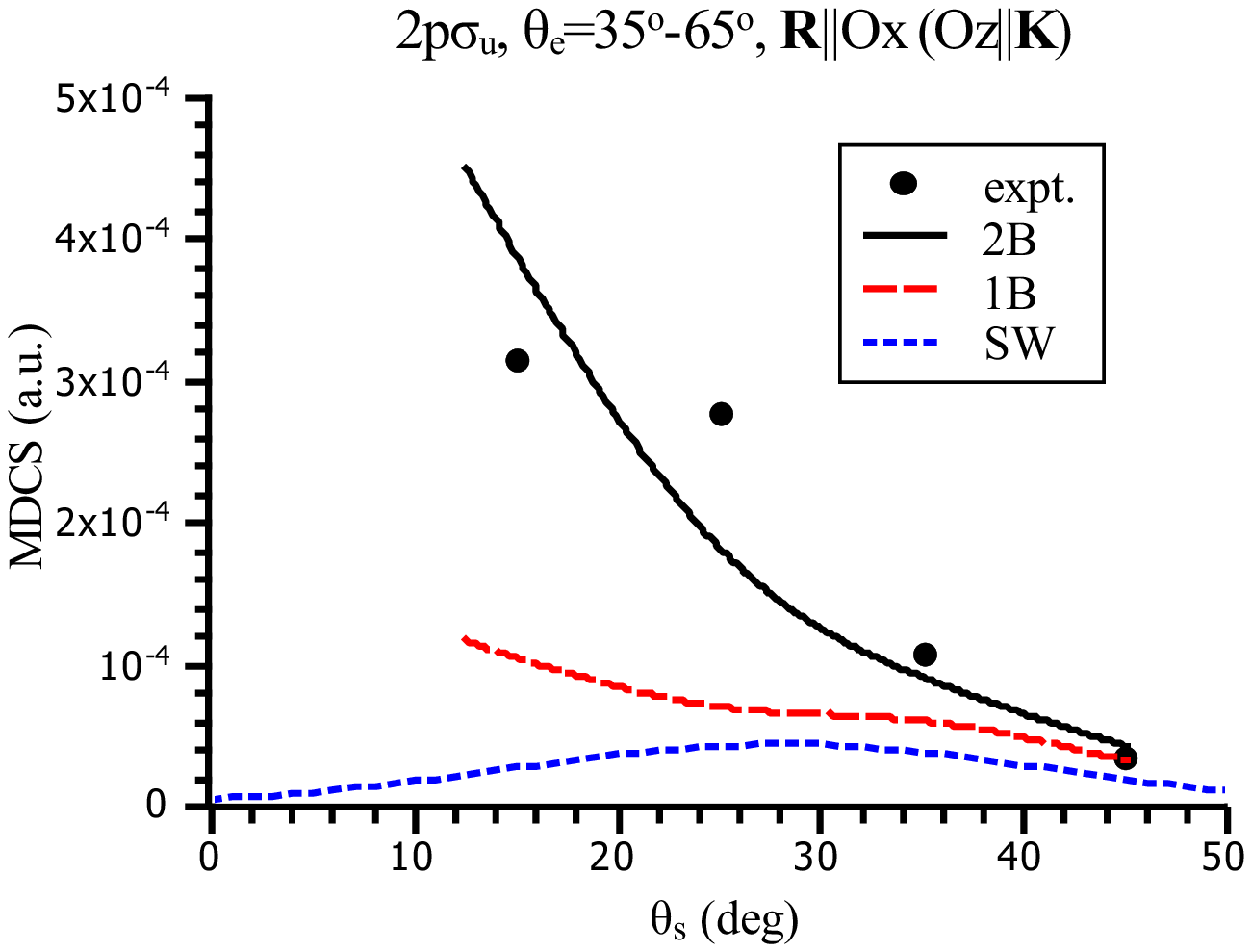}
\\(a)\\
\includegraphics[angle=-90,width=0.45\textwidth]{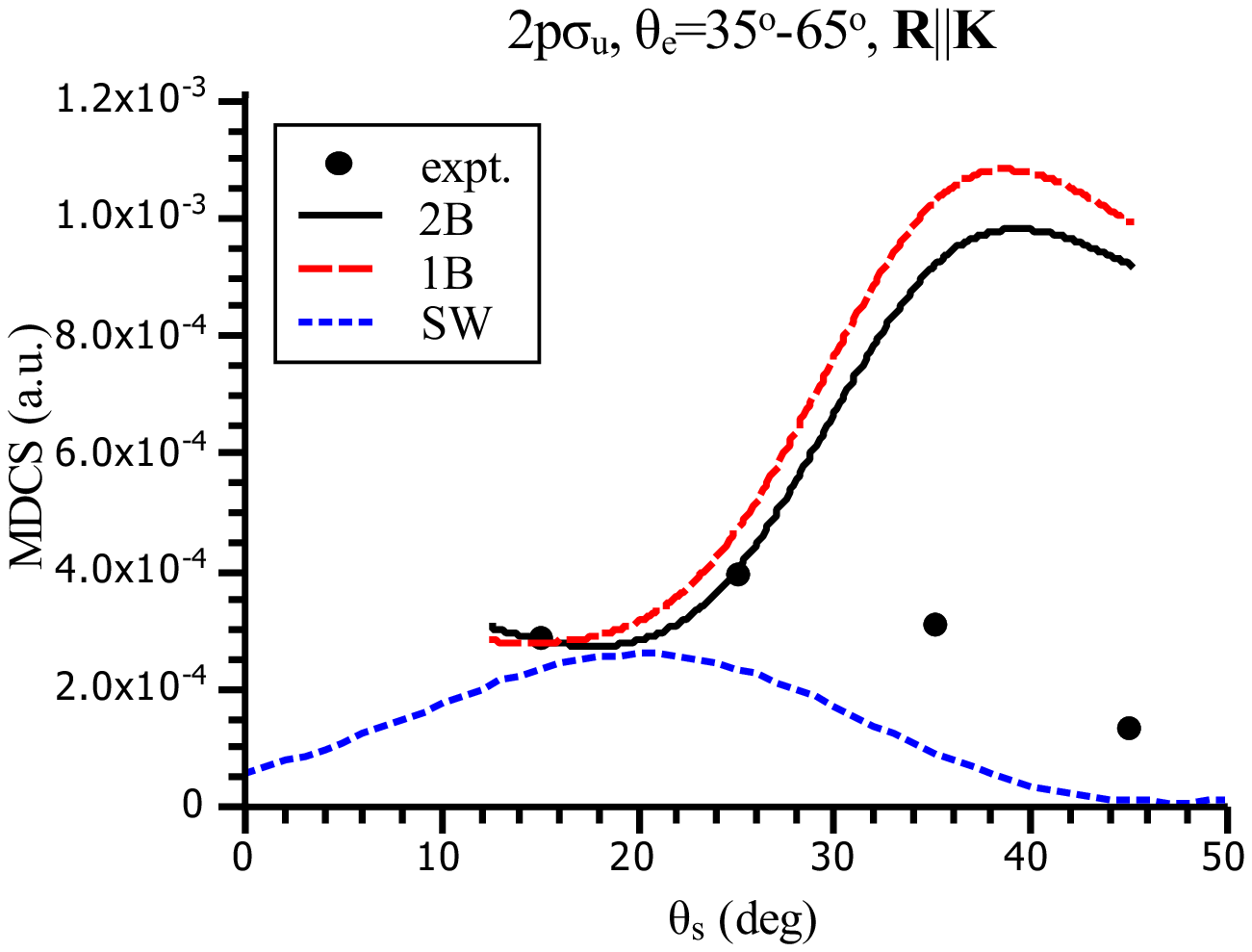}
\\(b)\\
\caption{(Color online) The same as on Fig.\ref{FIGsigma2psu}, but
for $Oz\parallel \mathbf{K}$.\label{FIGsigma2psuK}}
\end{figure}

We next consider the ionization-excitation of H$_2$ with formation
of the residual H$_2^+$ in $2p\sigma_u$ level. On Figures
\ref{FIGsigma2psu} and \ref{FIGsigma2psuK} the variations of the
MDCS for $2p\sigma_u$-case for the different orientations
described above are given. Here the results concerning the case of
perpendicular alignment to the incident plane designated by
$\mathbf{R}\parallel Oy$ are not presented as they are exactly
zero in all theoretical methods (due to symmetry of final state)
and are negligibly small in the experimental results.

The main observation that we can make from these figures is that
the results obtained by PSECS-1B and PSECS-2B do not reproduce the
structures of the experimental points. 
To go beyond  plane wave description for the incident and scattered waves we apply the PSECS-SW approach described above.
As mentioned there, this approach takes into account the
distortion undergone by the incident and the scattered waves due
to their interaction with the nuclei and due to the
post-collisional interaction with the ejected electron. It has
meanwhile the disadvantage of being unable to describe the
final-state scattering and the sequential mechanism. The
dominating term in the projection in equation (\ref{projPhi0}) of
the initial state on $2p\sigma_u$ orbital is of the same
$2p\sigma_u$ type, such that
\begin{eqnarray*}
\langle \varphi_{2p\sigma_u}| \Phi_0(\mathbf{r}_1,\mathbf{r}_2)\rangle\approx -0.0874\varphi_{2p\sigma_u}(\mathbf{r}_1).
\end{eqnarray*}

So the main idea here is to calculate the MDCS of the electron
impact ionization of H$_2^+$ (in the H$_2$ equilibrium
internuclear distance $R=$1.4 a.u.) for the $2p\sigma_u$ level,
and multiply it by the coefficient
$|c_{2p\sigma_u,2p\sigma_u}|^2=0.00764$. Since the ionization
energy of H$_2^+$($2p\sigma_u$) is less than that of the
ionization-excitation energy of H$_2$, we lowered the impact
energy to $E_i=160$ eV to keep the same energy values of the
ejected and scattered electrons in the ionization of H$_2$
($E_e=40$ eV and $E_s=100$ eV).  We should mention here, that the
SW method was not applied in the preceding cases for $2s\sigma_g$
and $2p\pi_u$, because of the necessity for
a large number of terms in the sum on right-hand-side of
Eq.(\ref{sumH2plus_channels}) in the case of the final ion state $2s\sigma_g$.

The PSECS-SW results are shown on Figures \ref{FIGsigma2psu} and
\ref{FIGsigma2psuK}. The corresponding PSECS-SW curves have the
same order of magnitude, but different structures compared to
those of the PSECS-1B and 2B curves. The dependence of PSECS-SW
results on $\theta_s$ seems to be closer to the experimental.

In what follows we will try to discuss the physical reasons of the
large discrepancy that exists between the PSECS-1B/2B and the
PSECS-SW results in the case of $2p\sigma_u$. It is easy to see,
that the most significant difference of PSECS-1B/2B from the
experimental data in both cases $2s\sigma_g+2p\pi_u$ and
$2p\sigma_u$ is observed for $\mathbf{R}\parallel\mathbf{K}$
(Figures \ref{FIGsigma2ssgK}b and \ref{FIGsigma2psuK}b), while for
$\mathbf{R}\perp\mathbf{K}$ (Figures \ref{FIGsigma2ssgK}a and
\ref{FIGsigma2psuK}a) the results of PSECS-1B/2B are close to the
experimental ones. In the case of $2p\sigma_u$, the large
difference from the experimental results is also observed for the
orientations parallel and perpendicular to the incidence direction
$\mathbf{R}\parallel\mathbf{k}_i$ and
$\mathbf{R}\perp\mathbf{k}_i$. This difference is smaller than
that of the case $\mathbf{R}\parallel\mathbf{K}$ and higher than
that of $\mathbf{R}\perp\mathbf{K}$. So, the angular domain of the
molecular orientation near that of the momentum transfer, where
PSECS-1B/2B are failing, is much wider for $2p\sigma_u$ than for
$2s\sigma_g+2p\pi_u$.

This observations can be explained using the known two-center
interference model \cite{Stia2003,Stia2002} of the impact
ionization of H$_2$. Although this model in its simplest form is
unable to give correct quantitative fitting of TDCS
\cite{Senftleben2}, it is useful for qualitative analysis of its
structure \cite{Serov2012}.
Let us consider for the case of impact ionization-excitation of H$_2$
with formation of H$_2^+$ in $2p\sigma_u$ state, the first Born
matrix element
\begin{eqnarray}
f_{\text{1B}}=-\frac{2}{K^2}\langle \mathbf{k}_e\, 2p\sigma_u|
\exp(i\mathbf{K}\cdot\mathbf{r}_1)+\exp(i\mathbf{K}\cdot\mathbf{r}_2)
| 0 \rangle.
\end{eqnarray}
Here $|0 \rangle\equiv\Phi_0(\mathbf{r}_1,\mathbf{r}_2)$ represents the ground state of H$_2$ and
\begin{eqnarray}
| \mathbf{k}_e\, 2p\sigma_u
\rangle=\frac{1}{\sqrt{2}}\left[\varphi_{\mathbf{k}_e}(\mathbf{r}_1)\varphi_{2p\sigma_u}(\mathbf{r}_2)+\varphi_{2p\sigma_u}(\mathbf{r}_1)\varphi_{\mathbf{k}_e}(\mathbf{r}_2)\right].
\end{eqnarray}
represents the final state, where
$\varphi_{2p\sigma_u}(\mathbf{r})$ is a wave function of
$2p\sigma_u$ of H$_2^+$ ion, and
$\varphi_{\mathbf{k}_e}(\mathbf{r})$ is a wave function of the
continuum state of H$_2^+$ ion. From the expansion
(\ref{expansionPhi0}) for $\Phi_0$ the projection $\langle
\mathbf{k}_e | 0 \rangle = 0$, because
$\varphi_{\mathbf{k}_e}(\mathbf{r})$ is orthogonal to all bound
states of H$_2^+$, and $\langle 2p\sigma_u| 0 \rangle \approx
c_{2p\sigma_u,2p\sigma_u} \varphi_{2p\sigma_u}(\mathbf{r}) $ as
mentioned earlier. So we can write
\begin{eqnarray}
f_{\text{1B}}\simeq -\frac{2^{3/2}}{K^2}c_{2p\sigma_u,2p\sigma_u}\langle
\mathbf{k}_e| \exp(i\mathbf{K}\cdot\mathbf{r})| 2p\sigma_u
\rangle.
\end{eqnarray}
If we assume, that the two centers of the molecule are so far,
that H$_2^+$ state can be presented by the combination of two
single-center atomic functions
\[ \varphi_{2p\sigma_u}(\mathbf{r}) \simeq \frac{1}{\sqrt{2}} [\varphi_{1s}(\mathbf{r}-\mathbf{R}/2)-\varphi_{1s}(\mathbf{r}+\mathbf{R}/2)],\]
and use a plane wave description for the final state
$\varphi_{\mathbf{k}_e}(\mathbf{r})\simeq
(2\pi)^{-3/2}\exp(i\mathbf{k}_e\cdot\mathbf{r})$, we can write
\begin{eqnarray}
f_{\text{1B}}\simeq
2c_{2p\sigma_u,2p\sigma_u}f_{\text{1B}1s}\sin[(\mathbf{K}-\mathbf{k}_e)\cdot\mathbf{R}/2].\label{interfer1B}
\end{eqnarray}
Here \begin{eqnarray} f_{\text{1B}1s}=-\frac{2}{K^2}\langle
\mathbf{k}_e | \exp(i\mathbf{K}\cdot\mathbf{r}) | 1s \rangle.
\end{eqnarray}
is the first-Born amplitude of the impact ionization of a single
hydrogen atom in the ground state. Eq.(\ref{interfer1B}) contains
an interference factor resulting from the two-center nature of the
target. Note that here the interference factor is
$\sin[(\mathbf{K}-\mathbf{k}_e)\cdot\mathbf{R}/2]$, and not
$\cos[(\mathbf{K}-\mathbf{k}_e)\cdot\mathbf{R}/2]$ as in
\cite{Stia2002}, because we consider the ungerade final state. The interference peak of the cross section in the case $\mathbf{R}\parallel\mathbf{K}$ is obtained for $KR=\pi$,
while for $\mathbf{R}\perp\mathbf{K}$ the interference factor does
not depend on $K$ and thus not on $\theta_s$. The positions of the
maxima of the PSECS-1B and PSECS-2B curves for
$\mathbf{R}\parallel\mathbf{K}$ on Fig.\ref{FIGsigma2psuK}b
confirms this result. For PSESC-SW the interference peak is
observed for lower values of the scattering angle.

Hence, the large discrepancy of PSECS-1B/2B results from the
experimental ones for the molecular orientation
$\mathbf{R}\parallel\mathbf{K}$ in the $2p\sigma_u$ case appears
to be a consequence of the strong sensibility of the positions of
the interference maxima to the deviation of the description of the
incident/scattered electron from the plane wave.  This is caused
by the interaction of the electron with the molecule before and
after the collision. In $2s\sigma_g$ and $2p\pi_u$ states, the
overlap between single-center functions is larger than that in
$2p\sigma_u$, and the two-center interference is less evident.
This can be the reason why the plane wave approximation for the
incident/scattered electron gives better results for these levels.


\section{Conclusion}

We have calculated the multifold differential cross section of
electron impact  dissociative ionization of H$_2$ by three
different variants of the exterior complex scaling procedure in
prolate spheroidal coordinates (PSECS). Our results are compared
to those of a recent experiment that detects in coincidence, for
the first time, the emerging electrons and one of the protons. In
the first variant, designated PSECS-1B, the two target electrons
are treated \textit{ab initio} while the interaction of the
incident-scattered electron is taken into account using the first
term of the Born series. In the second one, PSECS-2BCD, the second
Born term is introduced in the dipole approximation.  In the third
approach, designated PSECS-SW, applied to the
ionization-excitation to the $2p\sigma_u$ level of H$_2^+$, the
multi-configurational single active electron approximation is used
for the target, while the interaction of the incident electron
with target is described \textit{ab initio}. Our results obtained
by the PSECS-1B and 2B agree quite well with experimental results
concerning the ionization excitation to the $2s\sigma_g$ and
$2p\pi_u$ levels of H$_2^+$ and do not agree with those concerning
the $2p\sigma_u$ level. This can be explained by the fact, that
the the scattering angle values corresponding to the maxima are
very sensitive to the interaction of the incident electron with
the molecule, in situations where the maxima is caused by the
two-center interference. The PSECS-SW approach reproduces quite
well the structure of the experimental curve in this case.

\acknowledgments V.V.S. acknowledge Dr. T. Sergeeva for help. This work was partially supported by the Russian Foundation for Basic Research, grant No. 11-01-00523-a.

\end{document}